\documentclass[aps,prl,showpacs,twocolumn,groupedaddress]{revtex4}  
\usepackage{graphicx}  
\usepackage{dcolumn}   
\usepackage{bm}        
\usepackage{amssymb}   

\begin{document}



\hspace{5.2in} \mbox{FERMILAB-PUB-07/094-E}

\title{Measurement of the $\Lambda_b$ lifetime in the exclusive decay $\Lambda_b \rightarrow J/\psi \Lambda$}
%
\author{
V.M.~Abazov,$^{35}$ B.~Abbott,$^{75}$ M.~Abolins,$^{65}$ B.S.~Acharya,$^{28}$ M.~Adams,$^{51}$ T.~Adams,$^{49}$ E.~Aguilo,$^{5}$
S.H.~Ahn,$^{30}$ M.~Ahsan,$^{59}$ G.D.~Alexeev,$^{35}$
% ...
\\
\vskip 0.30cm
\centerline{(D\O\ Collaboration)}
\vskip 0.30cm
}
\affiliation{
\centerline{$^{1}$Universidad de Buenos Aires, Buenos Aires, Argentina}
\centerline{$^{2}$LAFEX, Centro Brasileiro de Pesquisas F{\'\i}sicas,
                  Rio de Janeiro, Brazil}
\centerline{$^{3}$Universidade do Estado do Rio de Janeiro,
                  Rio de Janeiro, Brazil}
\centerline{$^{4}$Instituto de F\'{\i}sica Te\'orica, Universidade
                  Estadual Paulista, S\~ao Paulo, Brazil}
\centerline{$^{5}$University of Alberta, Edmonton, Alberta, Canada,
                  Simon Fraser University, Burnaby, British Columbia, Canada,}
\centerline{York University, Toronto, Ontario, Canada, and
                  McGill University, Montreal, Quebec, Canada}
\centerline{$^{6}$University of Science and Technology of China, Hefei,
                  People's Republic of China}
\centerline{$^{7}$Universidad de los Andes, Bogot\'{a}, Colombia}
\centerline{$^{8}$Center for Particle Physics, Charles University,
                  Prague, Czech Republic}
\centerline{$^{9}$Czech Technical University, Prague, Czech Republic}
\centerline{$^{10}$Center for Particle Physics, Institute of Physics,
                   Academy of Sciences of the Czech Republic,
                   Prague, Czech Republic}
\centerline{$^{11}$Universidad San Francisco de Quito, Quito, Ecuador}
\centerline{$^{12}$Laboratoire de Physique Corpusculaire, IN2P3-CNRS,
                   Universit\'e Blaise Pascal, Clermont-Ferrand, France}
\centerline{$^{13}$Laboratoire de Physique Subatomique et de Cosmologie,
                   IN2P3-CNRS, Universite de Grenoble 1, Grenoble, France}
\centerline{$^{14}$CPPM, IN2P3-CNRS, Universit\'e de la M\'editerran\'ee,
                   Marseille, France}
\centerline{$^{15}$Laboratoire de l'Acc\'el\'erateur Lin\'eaire,
                   IN2P3-CNRS et Universit\'e Paris-Sud, Orsay, France}
\centerline{$^{16}$LPNHE, IN2P3-CNRS, Universit\'es Paris VI and VII,
                   Paris, France}
\centerline{$^{17}$DAPNIA/Service de Physique des Particules, CEA, Saclay,
                   France}
\centerline{$^{18}$IPHC, Universit\'e Louis Pasteur et Universit\'e
                   de Haute Alsace, CNRS, IN2P3, Strasbourg, France}
\centerline{$^{19}$IPNL, Universit\'e Lyon 1, CNRS/IN2P3, Villeurbanne, France
                   and Universit\'e de Lyon, Lyon, France}
\centerline{$^{20}$III. Physikalisches Institut A, RWTH Aachen,
                   Aachen, Germany}
\centerline{$^{21}$Physikalisches Institut, Universit{\"a}t Bonn,
                   Bonn, Germany}
\centerline{$^{22}$Physikalisches Institut, Universit{\"a}t Freiburg,
                   Freiburg, Germany}
\centerline{$^{23}$Institut f{\"u}r Physik, Universit{\"a}t Mainz,
                   Mainz, Germany}
\centerline{$^{24}$Ludwig-Maximilians-Universit{\"a}t M{\"u}nchen,
                   M{\"u}nchen, Germany}
\centerline{$^{25}$Fachbereich Physik, University of Wuppertal,
                   Wuppertal, Germany}
\centerline{$^{26}$Panjab University, Chandigarh, India}
\centerline{$^{27}$Delhi University, Delhi, India}
\centerline{$^{28}$Tata Institute of Fundamental Research, Mumbai, India}
\centerline{$^{29}$University College Dublin, Dublin, Ireland}
\centerline{$^{30}$Korea Detector Laboratory, Korea University,
                   Seoul, Korea}
\centerline{$^{31}$SungKyunKwan University, Suwon, Korea}
\centerline{$^{32}$CINVESTAV, Mexico City, Mexico}
\centerline{$^{33}$FOM-Institute NIKHEF and University of
                   Amsterdam/NIKHEF, Amsterdam, The Netherlands}
\centerline{$^{34}$Radboud University Nijmegen/NIKHEF, Nijmegen, The
                  Netherlands}
\centerline{$^{35}$Joint Institute for Nuclear Research, Dubna, Russia}
\centerline{$^{36}$Institute for Theoretical and Experimental Physics,
                   Moscow, Russia}
\centerline{$^{37}$Moscow State University, Moscow, Russia}
\centerline{$^{38}$Institute for High Energy Physics, Protvino, Russia}
\centerline{$^{39}$Petersburg Nuclear Physics Institute,
                   St. Petersburg, Russia}
\centerline{$^{40}$Lund University, Lund, Sweden, Royal Institute of
                   Technology and Stockholm University, Stockholm,
                   Sweden, and}
\centerline{Uppsala University, Uppsala, Sweden}
\centerline{$^{41}$Physik Institut der Universit{\"a}t Z{\"u}rich,
                   Z{\"u}rich, Switzerland}
\centerline{$^{42}$Lancaster University, Lancaster, United Kingdom}
\centerline{$^{43}$Imperial College, London, United Kingdom}
\centerline{$^{44}$University of Manchester, Manchester, United Kingdom}
\centerline{$^{45}$University of Arizona, Tucson, Arizona 85721, USA}
\centerline{$^{46}$Lawrence Berkeley National Laboratory and University of
                   California, Berkeley, California 94720, USA}
\centerline{$^{47}$California State University, Fresno, California 93740, USA}
\centerline{$^{48}$University of California, Riverside, California 92521, USA}
\centerline{$^{49}$Florida State University, Tallahassee, Florida 32306, USA}
\centerline{$^{50}$Fermi National Accelerator Laboratory,
            Batavia, Illinois 60510, USA}
\centerline{$^{51}$University of Illinois at Chicago,
            Chicago, Illinois 60607, USA}
\centerline{$^{52}$Northern Illinois University, DeKalb, Illinois 60115, USA}
\centerline{$^{53}$Northwestern University, Evanston, Illinois 60208, USA}
\centerline{$^{54}$Indiana University, Bloomington, Indiana 47405, USA}
\centerline{$^{55}$University of Notre Dame, Notre Dame, Indiana 46556, USA}
\centerline{$^{56}$Purdue University Calumet, Hammond, Indiana 46323, USA}
\centerline{$^{57}$Iowa State University, Ames, Iowa 50011, USA}
\centerline{$^{58}$University of Kansas, Lawrence, Kansas 66045, USA}
\centerline{$^{59}$Kansas State University, Manhattan, Kansas 66506, USA}
\centerline{$^{60}$Louisiana Tech University, Ruston, Louisiana 71272, USA}
\centerline{$^{61}$University of Maryland, College Park, Maryland 20742, USA}
\centerline{$^{62}$Boston University, Boston, Massachusetts 02215, USA}
\centerline{$^{63}$Northeastern University, Boston, Massachusetts 02115, USA}
\centerline{$^{64}$University of Michigan, Ann Arbor, Michigan 48109, USA}
\centerline{$^{65}$Michigan State University,
            East Lansing, Michigan 48824, USA}
\centerline{$^{66}$University of Mississippi,
            University, Mississippi 38677, USA}
\centerline{$^{67}$University of Nebraska, Lincoln, Nebraska 68588, USA}
\centerline{$^{68}$Princeton University, Princeton, New Jersey 08544, USA}
\centerline{$^{69}$State University of New York, Buffalo, New York 14260, USA}
\centerline{$^{70}$Columbia University, New York, New York 10027, USA}
\centerline{$^{71}$University of Rochester, Rochester, New York 14627, USA}
\centerline{$^{72}$State University of New York,
            Stony Brook, New York 11794, USA}
\centerline{$^{73}$Brookhaven National Laboratory, Upton, New York 11973, USA}
\centerline{$^{74}$Langston University, Langston, Oklahoma 73050, USA}
\centerline{$^{75}$University of Oklahoma, Norman, Oklahoma 73019, USA}
\centerline{$^{76}$Oklahoma State University, Stillwater, Oklahoma 74078, USA}
\centerline{$^{77}$Brown University, Providence, Rhode Island 02912, USA}
\centerline{$^{78}$University of Texas, Arlington, Texas 76019, USA}
\centerline{$^{79}$Southern Methodist University, Dallas, Texas 75275, USA}
\centerline{$^{80}$Rice University, Houston, Texas 77005, USA}
\centerline{$^{81}$University of Virginia, Charlottesville,
            Virginia 22901, USA}
\centerline{$^{82}$University of Washington, Seattle, Washington 98195, USA}
}
\date{April 30, 2007}

\begin{abstract}

We have measured the $\Lambda_{b}$ lifetime using the exclusive decay $\Lambda_b \to J/\psi\mbox{ }\Lambda$, based on 1.2 fb$^{-1}$ of data
collected with the D0 detector during 2002--2006. From 171 reconstructed $\Lambda_{b}$ decays, where the $J/\psi$ and $\Lambda$ are identified
via the decays $J/\psi \to \mu^{+}\mu^{-}$ and $\Lambda \to p\pi$, we measured the $\Lambda_{b}$ lifetime to be $\tau(\Lambda_b) =
1.218^{+0.130}_{-0.115}\mbox{(stat)}\pm 0.042 \mbox{(syst)}$ ps. We also measured the $B^{0}$ lifetime in the decay $B^0 \to
J/\psi(\mu^{+}\mu^{-})K^{0}_{S}(\pi^+ \pi^-)$ to be $\tau(B^0) = 1.501^{+0.078}_{-0.074}\mbox{(stat)}\pm 0.050 \mbox{(syst)}$ ps, yielding a
lifetime ratio of $\tau(\Lambda_b)/\tau(B^0) = 0.811^{+0.096}_{-0.087}\mbox{(stat)}\pm 0.034\mbox{(syst)}$. 

\end{abstract}

\maketitle

\vskip.5cm

\newpage

Lifetime measurements of $b$ hadrons provide important information on the interactions between heavy and light quarks. At leading order in Heavy
Quark Effective Theory (HQET)~\cite{HQET}, light quarks are considered spectators and all $b$ hadrons have the same lifetime. Differences arise
at higher orders when corrections from interactions are taken into account. For HQET calculations of order $1/m_{b}^2$, where $m_{b}$ is the
mass of the $b$ quark, the agreement between the predicted lifetimes and the experimental results is excellent for $B$ mesons~\cite{Neubert}.
However, in the $b$ baryon sector, the world average of measurements of $\tau(\Lambda_{b})/\tau(B^{0}) = 0.844\pm 0.043$~\cite{HFAG2006} is
smaller than the prediction of the ratio at this order.
Recently there have been significant improvements in theoretical calculations of $\tau(\Lambda_{b})/\tau(B^{0})$. Next-to-leading order effects
in QCD~\cite{Franco}, corrections at ${\cal{O}}(1/m_{b}^{4})$ in HQET~\cite{Gabbiani}, and lattice QCD studies~\cite{DiPierro}, have led to an
improved theoretical prediction, $\tau(\Lambda_{b})/\tau(B^{0})=0.88\pm0.05$~\cite{Tarantino}. This value agrees with previous experiments to
within the current theoretical and experimental uncertainties. However, a recent precise measurement~\cite{CDFLb} reports a value of the $\Lambda_{b}$
lifetime consistent with $B$ meson lifetimes, and the ratio $\tau(\Lambda_{b})/\tau(B^{0})$ consistent with unity. Additional precise
measurements of the $\Lambda_{b}$ lifetime and $\tau(\Lambda_{b})/\tau(B^{0})$ ratio may help settle this question.


In this Letter, we report measurements of the $\Lambda_{b}$ lifetime using the exclusive decay  $\Lambda_{b}\rightarrow J/\psi +\Lambda$, and
its ratio to the $B^0$ lifetime using the $B^0\rightarrow J/\psi +K^{0}_{S}$ decay channel. This $B^0$ decay channel is chosen because of its
similar topology to the $\Lambda_b$ decay.  The $J/\psi$ is reconstructed from the $\mu^+ \mu^-$ decay mode, the $\Lambda$ from $p\pi^{-}$, and
the $K^0_S$ from $\pi^+ \pi^-$. Throughout this Letter, the appearance of a specific charge state also implies its charge conjugate. The data
used in this analysis were collected during 2002--2006 with the D0 detector in Run II of the Tevatron Collider at a center-of-mass energy of
1.96~TeV, and correspond to an integrated luminosity of 1.2~fb$^{-1}$.

The D0 detector is described in detail elsewhere~\cite{run2det}. The
detector components most relevant to this analysis are the central
tracking and the muon systems. The former consists of a silicon
microstrip tracker (SMT) and a central scintillating fiber tracker
(CFT) surrounded by a 2~T superconducting solenoidal magnet. The SMT
has a design optimized for tracking and vertexing for pseudorapidity
of $|\eta|<3$~\cite{etadef}. For charged particles, the resolution
on the distance of closest approach as provided by the tracking
system is approximately 50~$\mu$m for tracks with $p_T \approx
1$~GeV/$c$, where $p_T$ is the component of the momentum
perpendicular to the beam axis. It improves asymptotically to
15~$\mu$m for tracks with $p_T > 10$~GeV/$c$. Preshower detectors
and electromagnetic and hadronic calorimeters surround the tracker.
The muon system is located beyond the calorimeter, and consists of
multilayer drift chambers and scintillation counters inside 1.8~T
toroidal magnets, and two similar layers outside the toroids. Muon
identification for $|\eta|<1$ relies on 10 cm wide drift tubes,
while 1 cm mini-drift tubes are used for $1<|\eta|<2$.

The primary vertex of the $p\bar{p}$ interaction is determined for
each event using the average position of the beam-collision in the
plane perpendicular to the beam as a constraint. The precision of
the primary vertex reconstruction is on average 20~$\mu m$ in the
plane perpendicular to the beam and about 40~$\mu m$ along the
direction of the beam.

We base our data selection on reconstructed charged tracks and identified
muons. Although we do not require any specific trigger, most of the selected
events satisfy dimuon or muon triggers. To avoid a trigger bias in the lifetime
measurement, we reject events that depend on impact parameter based triggers.
We start the $\Lambda_{b}$ and $B^{0}$ reconstruction by searching for events
with $J/\psi$ mesons. We then search in these events for $\Lambda$ and
$K^{0}_{S}$ particles. To reconstruct $J/\psi\to\mu^{+}\mu^{-}$ candidates, we
select events with at least two muons of opposite charge reconstructed in the
tracker and the muon system. The track of each muon candidate must either match
hits in the muon system, or have calorimeter energies consistent with a
minimum-ionizing particle along the direction of hits extrapolated from the
tracking layers. For at least one of the muons, we require hits in all three
layers of the muon detector. Both muons are required to have $p_T>$2.5~GeV/$c$
if they are in the region $|\eta|<$1. The muon tracks are constrained to
originate from a common vertex with a $\chi^2$ probability greater than 1\%,
and each $J/\psi$ candidate is required to have a mass in the range
2.80--3.35~GeV/$c^2$. The $\Lambda \to p \pi^-$ decays are reconstructed from
two tracks of opposite charge constrained to a common vertex with a $\chi^2$
probability greater than 1\%. Each $\Lambda$ candidate is required to have a
mass in the range 1.100--1.128~GeV/$c^2$. The proton mass is assigned to the
track of higher $p_{T}$, as observed in Monte Carlo studies. To suppress
contamination from cascade decays of more massive baryons such as
$\Sigma^{0}\to\Lambda\gamma$ or $\Xi^{0}\to\Lambda\pi^{0}$, we require the
cosine of the angle between the $\bm{p}_T$ vector of the $\Lambda$ and the
vector in the perpendicular plane from the $J/\psi$ vertex to the $\Lambda$
decay vertex to be larger than 0.9999. For $\Lambda$'s that decay from
$\Lambda_b$ the cosine of this angle is very close to one. The $K^0_S \to \pi^+
\pi^-$ selection follows the same criteria, except that for the $K^0_S$, the
mass window is 0.460--0.525~GeV/$c^2$, and pion mass assignments are used.

We reconstruct the $\Lambda_b$ and $B^0$ by performing a constrained fit to a common vertex for either the $\Lambda$ or $K^0_S$ and the two muon
tracks, with the latter constrained to the $J/\psi$ mass of 3.097~GeV/$c^2$~\cite{HFAG2006}. Because of their long decay lengths, a significant
fraction of $\Lambda$ and $K^0_S$ particles will decay outside the SMT. There is therefore no requirement of SMT hits on the tracks from
$\Lambda$ and $K^0_S$ decays. To reconstruct the $\Lambda_b$ ($B^0$), we first find the $\Lambda$ ($K^0_S$) decay vertex, and then extrapolate
the momentum vector of the ensuing particle and form a vertex together with the two muon tracks belonging to the $J/\psi$. If more than one
candidate is found in the event, the candidate with the best $\chi^2$ probability is selected as the $\Lambda_b$ ($B^0$) candidate. The mass is
required to be within the range 5.1--6.1~GeV/$c^2$ for $\Lambda_b$ candidates and within 4.9--5.7~GeV/$c^2$ for $B^0$ candidates. For the choice
of the final selection criteria, we optimize $S/\sqrt{S+B}$, where $S$ and $B$ are the number of signal ($\Lambda_{b}$) and background
candidates, respectively, by using Monte Carlo estimates for $S$ and data for $B$. For the Monte Carlo, we use {\sc pythia}~\cite{Pythia} and
{\sc evtgen}~\cite{EvtGen} to produce and decay particles, respectively, and {\sc geant3}~\cite{Geant} to simulate detector effects.
As a result of this optimization, the $p_T$ of the $\Lambda$ ($K^{0}_{S}$) is required to be greater than 2.4 (1.8)~GeV/$c$, and the total
momentum for both $\Lambda_{b}$ and $B^{0}$ is required to be greater than 5~GeV$/c$. Finally, any candidate which has been identified as a
$\Lambda_{b}$ is removed from the $B^{0}$ sample.


We determine the decay time of a $\Lambda_b$ or $B^0$ by measuring the distance traveled by the $b$ hadron candidate in a plane transverse to
the beam direction, and then applying a correction for the Lorentz boost. We define the transverse decay length as $L_{xy} = \bm{L}_{xy} \cdot
\bm{p}_T/p_T$ where $\bm{L}_{xy}$ is the vector that points from the primary vertex to the $b$ hadron decay vertex and $\bm{p}_T$ is the
transverse momentum vector of the $b$ hadron. The event-by-event value of the proper transverse decay length, $\lambda$, for the $b$ hadron
candidate is given by:
\begin{equation}
    \label{ctau} \lambda =\frac{L_{xy}}{(\beta\gamma)_{T}^{B}} = L_{xy}\, \frac{c M_{B}}{p_{T}},
\end{equation}
\noindent where $(\beta\gamma)_{T}^{B}$ and $M_{B}$ are the
transverse boost and the mass of the $b$ hadron.  In our
measurement, the value of $M_{B}$ in Eq.~\ref{ctau}\ is set to the
Particle Data Group (PDG) mass value of $\Lambda_{b}$ or
$B^0$~\cite{HFAG2006}. We require the uncertainty on $\lambda$ to be
less than 500~$\mu$m.


We perform a simultaneous unbinned maximum likelihood fit to the mass and proper decay length distributions.
The likelihood function ${\cal{L}}$ is defined by:
\begin{equation}
    \begin{array}{lcl}
        {\cal{L}} & = & \frac{(n_{s}+n_{b})^{N}}{N!}\exp\left(-n_{s}-n_{b}\right)\times\\
        & &\prod^{N}_{j=1} \left[\frac{n_{s}}{n_{s} + n_{b}}{\cal{F}}_{\text{sig}}^{j}+ \frac{n_{b}}{n_{s} +
        n_{b}}{\cal{F}}_{\text{bkg}}^{j}\right],
    \end{array}
\end{equation}
where $n_{s}$ and $n_{b}$ are the expected number of signal and background events in the sample, respectively. $N$ is the total number of
events. ${\cal{F}}_{\text{sig}}^{j}$ (${\cal{F}}_{\text{bkg}}^{j}$) is the product of three probability density functions that model the mass,
proper decay length, and uncertainty on proper decay length distributions for the signal (background). We divide the background into two
categories, prompt and non-prompt. The prompt background is primarily due to direct production of $J/\psi$'s which are then randomly combined
with a $\Lambda$ or $K^0_S$ candidate in the event. The non-prompt background is mainly produced by the combination of $J/\psi$ mesons from $b$
hadron decays with $\Lambda$ or $K^0_S$ candidates present in the event.

For the signal, the mass distribution is modeled by a Gaussian function, and the $\lambda$ distribution is parametrized by an exponential decay
convoluted with the resolution function:
\begin{equation}
    G(\lambda_{j},\sigma_{j}) = \frac{1}{\sqrt{2\pi}s\sigma_{j}} \exp \left[\frac{-\lambda_{j}^{2}}{2(s\sigma_j)^{2}}\right],
    \label{res}
\end{equation}
where $\lambda_j$ and $\sigma_j$ represent $\lambda$ and its uncertainty, respectively, for a given decay $j$, and $s$ is a common scale
parameter introduced in the fit to account for a possible mis-estimate of $\sigma_j$. The convolution is defined by:
\begin{equation}
    S_{\lambda}(\lambda_{j},\sigma_{j})=\frac{1}{\lambda_{B}}\int^{\infty}_{0}G(x-\lambda_{j},\sigma_{j})\exp \left(\frac{-x}{\lambda_{B}}\right)
    dx,
\end{equation}
where $\lambda_{B}=c\tau_{B}$, and $\tau_{B}$ is the lifetime of the
$\Lambda_{b}$ ($B^{0}$). The distribution of the uncertainty of
$\lambda$ is modeled by an exponential function convoluted by a
Gaussian.

For the background, the mass distribution of the prompt component is
assumed to follow a flat distribution as observed in data when a cut
of $\lambda>$100 $\mu$m is applied. The non-prompt component is
modeled with a second-order polynomial function. The $\lambda$
distribution is parametrized by the resolution function for the
prompt component, and by the sum of negative and positive
exponential functions for the non-prompt component. A positive and a
negative exponential functions model combinatorial background, and
an exponential function accounts for long-lived heavy flavor decays.
The distribution of the uncertainty of $\lambda$ is modeled by two
exponential functions convoluted by a Gaussian.

\begin{figure*}
    \begin{tabular}{cc}
        \includegraphics[scale=.43]{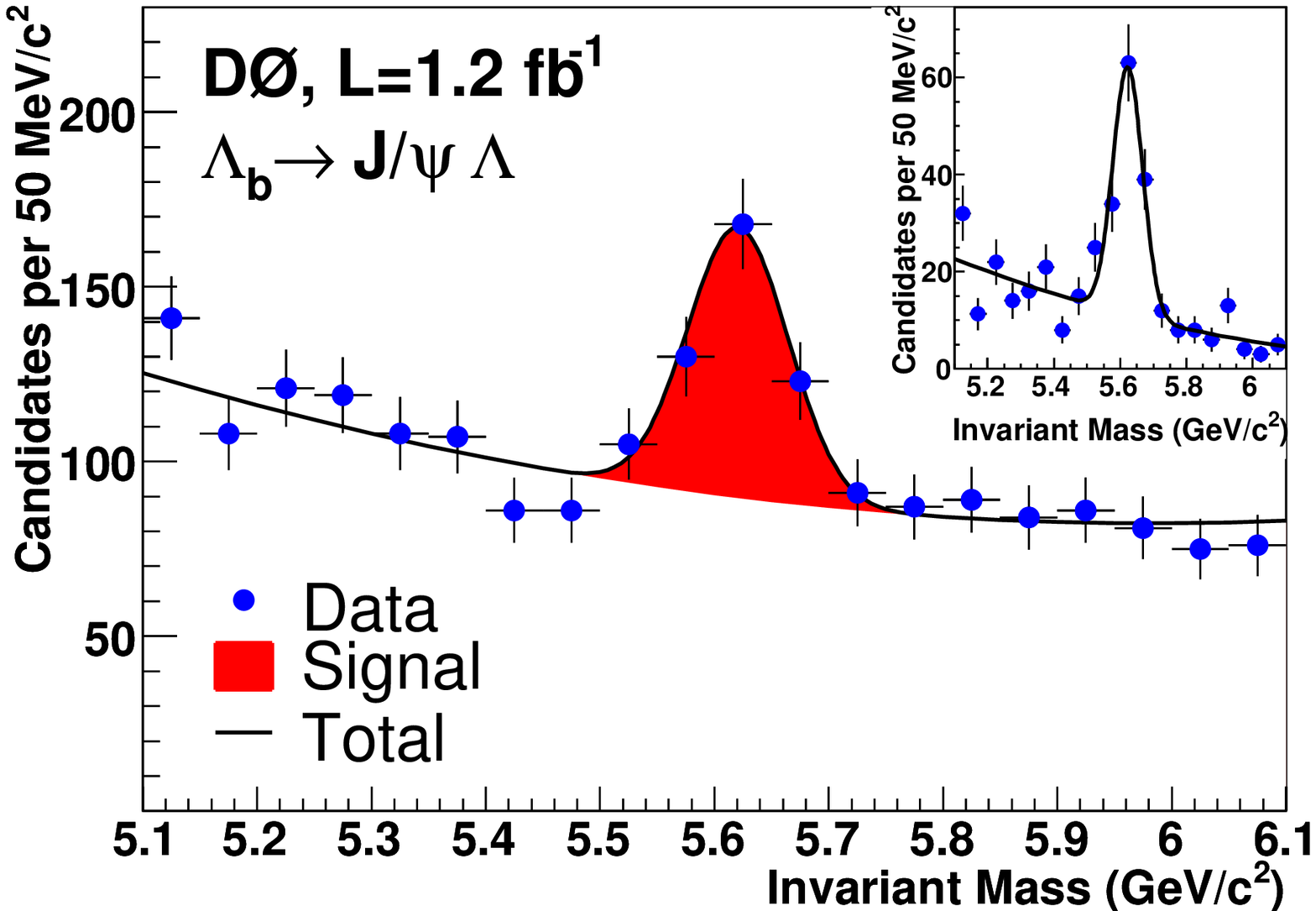} &
        \includegraphics[scale=.43]{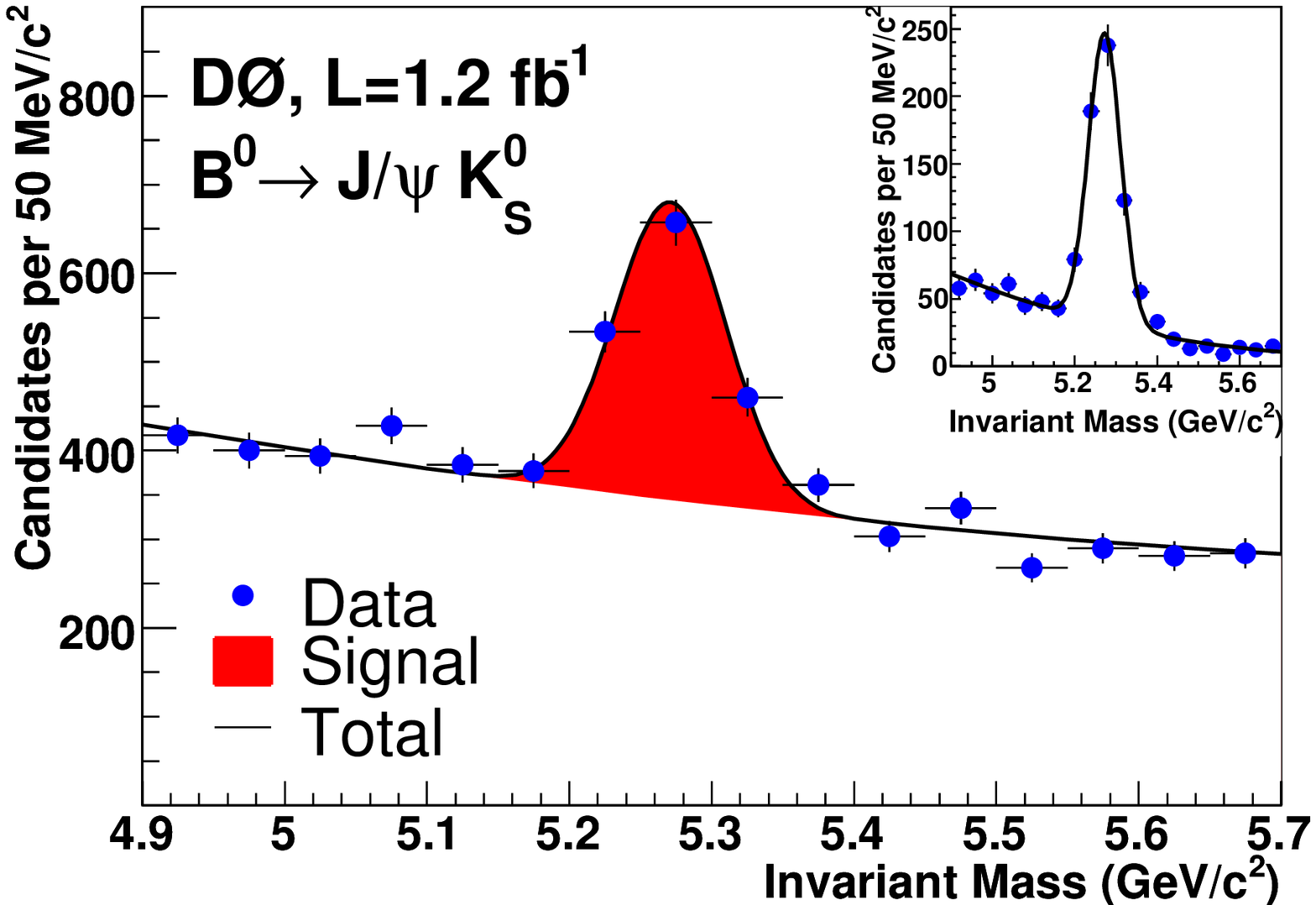}
    \end{tabular}
    \caption{\label{fig:lbm} Invariant mass distribution for $\Lambda_{b}$(left) and $B^{0}$(right)
    candidates, with the fit results superimposed. 
The inserts show the mass distributions after requiring $\lambda/\sigma>5$.
}
\end{figure*}

\begin{figure*}
    \begin{tabular}{cc}
        \includegraphics[scale=.43]{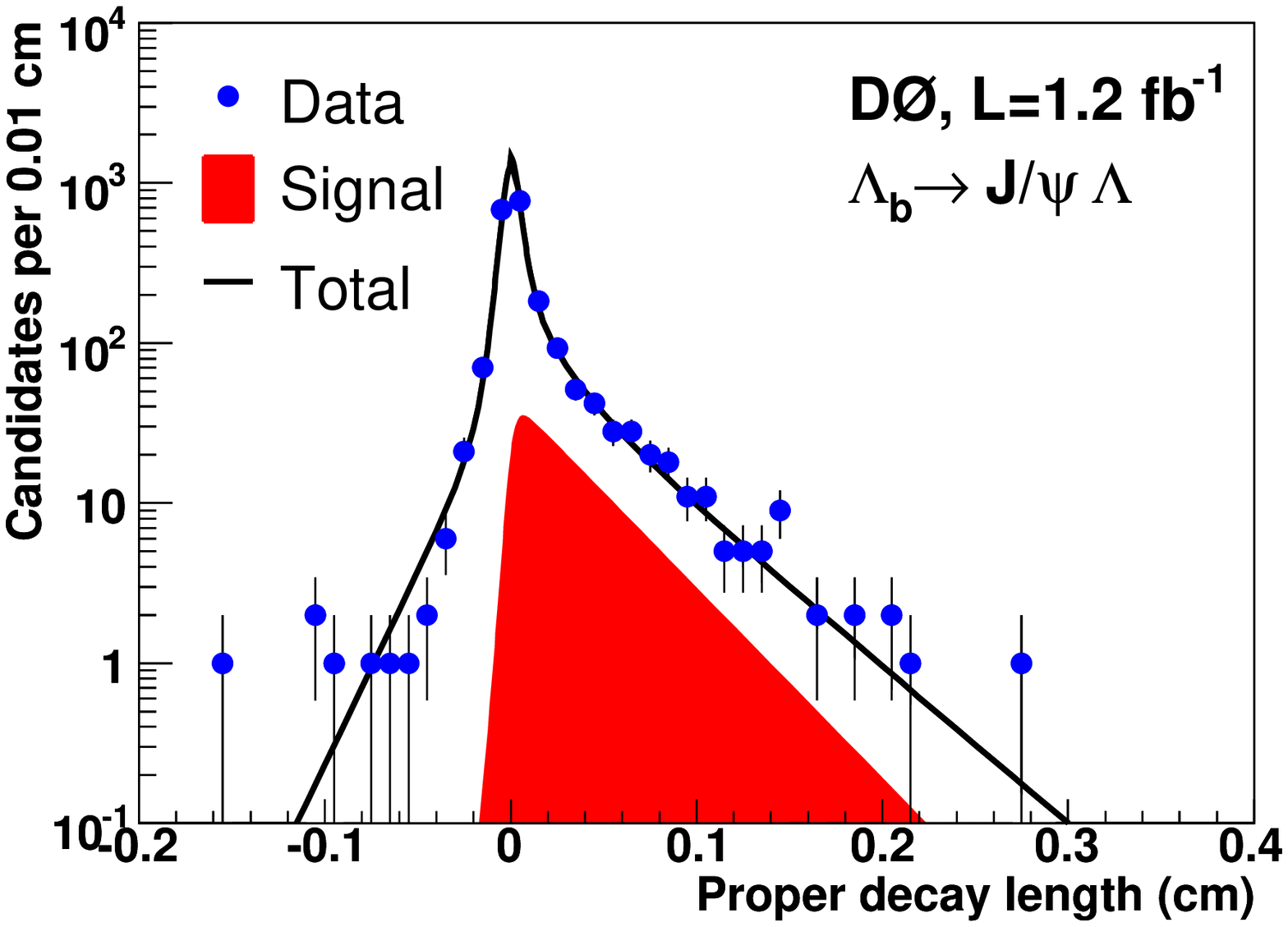} &
        \includegraphics[scale=.43]{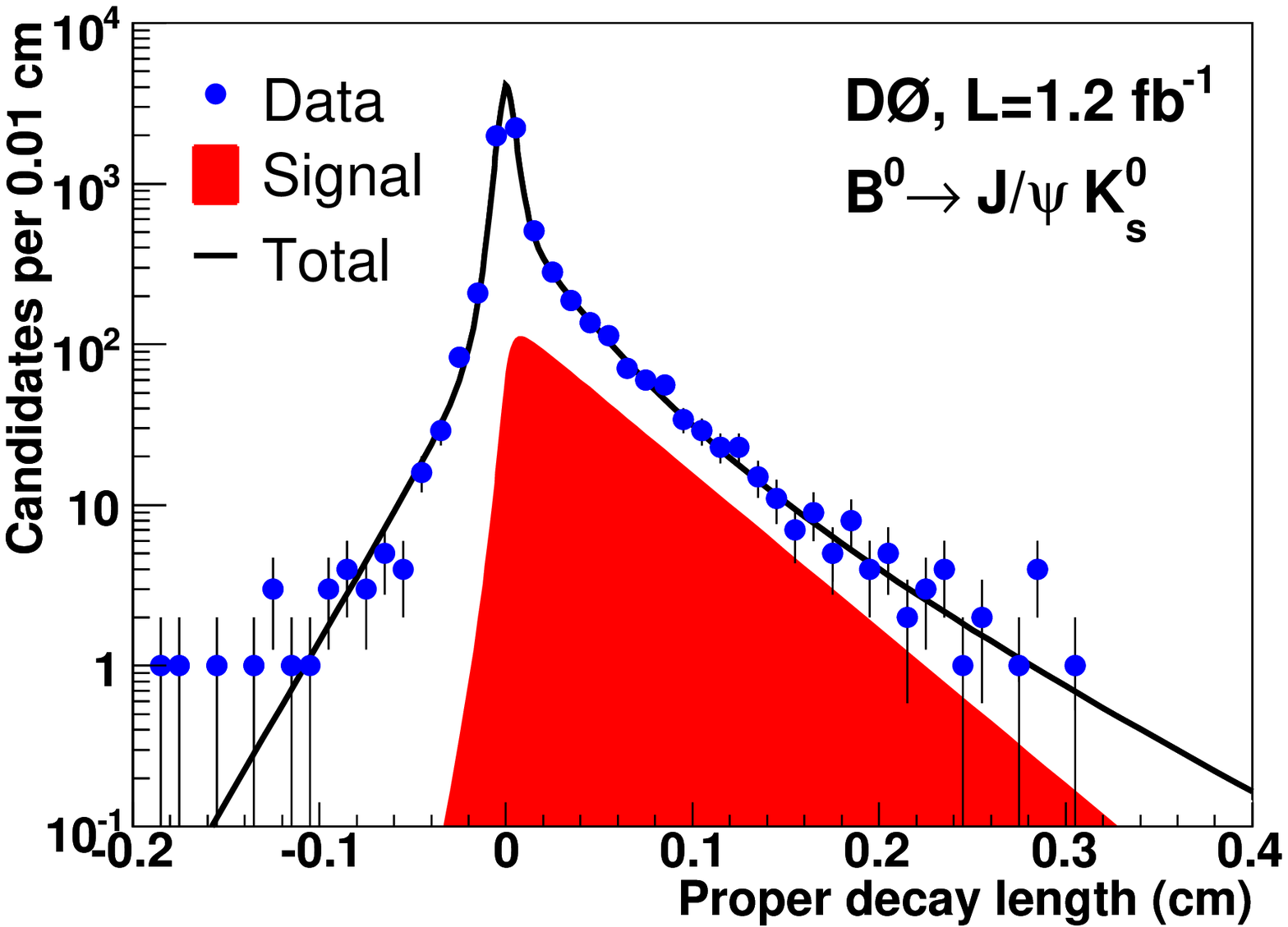}
    \end{tabular}
    \caption{\label{fig:lblt} Proper decay length distribution for $\Lambda_b$(left) and $B^{0}$(right)
    candidates, with the fit results superimposed. The shaded region represents the signal.}
\end{figure*}

We minimize $-2\ln{{\cal L}}$ to extract: $c\tau (\Lambda_b)=365.1^{+39.1}_{-34.7}$~$\mu$m and $c\tau(B^0)=450.0^{+23.5}_{-22.1}$~$\mu$m. From
the fits, we obtain $s=1.41\pm 0.05$ for the $\Lambda_b$ and $s=1.41\pm 0.03$ for the $B^0$. The numbers of signal
decays are $171\pm20$ $\Lambda_b$ and $717\pm38$ $B^0$. Figures~\ref{fig:lbm} and \ref{fig:lblt} show the mass and $\lambda$ distributions for
the $\Lambda_b$ and $B^{0}$ candidates. Fit results are superimposed.
Table~\ref{tab:table2} summarizes the systematic uncertainties in our measurements. The contribution from possible misalignment of the SMT
detector was estimate to be 5.4~$\mu$m~\cite{PRL94-102001}. We estimate the systematic uncertainty due to the models for the $\lambda$ and mass
distributions by varying the parametrizations of the different components: (i) the resolution function is modeled by two Gaussian functions
instead of one, (ii) the exponential functions in the non-prompt background are replaced by exponentials convoluted with the resolution
function, (iii) a uniform background is added to account for outlier events (this has only a negligible effect), (iv) the positive and negative
exponentials describing combinatorial non-prompt background are assumed to be symmetric, and (v) for the mass distribution of the non-prompt
background, a linear function is used instead of the nominal quadratic form. To take into account correlations between the effects of the
different models, a fit that combines all different model changes is performed. We quote the difference between the result of this fit and the
nominal fit as the systematic uncertainty.

The lifetime of the background events under the
$\Lambda_{b}$($B^{0}$) signal is mostly modeled by events in the low
and high mass sideband regions with respect to the peak. To estimate
the effect of any difference between the lifetime distributions of
these two regions, we perform separate fits to the $\Lambda_{b}$
($B^{0}$) mass regions of 5.1--5.8 and 5.4--6.1~GeV/$c^2$ (4.9--5.45
and 5.1--5.7~GeV/$c^2$) where the contributions from high and low
mass background events are reduced, respectively. The largest
difference between these fits and the nominal fit is quoted as the
systematic uncertainty due to this source.

We also study the contamination of the $\Lambda_b$ sample by $B^0$ events that pass the $\Lambda_b$ selection. From Monte Carlo studies, we
estimate that 6.5\% of $B^0$ events pass the $\Lambda_b$ selection criteria. However, the invariant mass of $B^{0}$ events which contaminate the
$\Lambda_{b}$ sample is distributed almost uniformly across the entire $\Lambda_{b}$ mass range, and their proper decay lengths therefore tend
to be incorporated in the long-lived component of the background.  To estimate the effect due to this contamination, we remove from the
$\Lambda_{b}$ sample any event which also passes the $B^{0}$ selection criteria, and we perform a fit to the remaining events. The difference
between this and the nominal fit is quoted as the systematic uncertainty due to the contamination. For the $B^{0}$, we do not consider this
source of systematic uncertainty since any event identified as $\Lambda_{b}$ is removed from the $B^{0}$ sample.

We perform several cross-checks of the lifetime measurements. The $J/\psi$ vertex is used instead of the $b$ hadron vertex, the mass windows are
varied, the reconstructed $b$ hadron mass is used instead of the PDG~\cite{HFAG2006} value, and the sample is split into different
pseudorapidity regions and different regions of azimuth. All results obtained with these variations are consistent with our measurement. We also
cross-check the fitting procedure and selection criteria by measuring the $\Lambda_{b}$ lifetime in Monte Carlo events. The lifetime obtained
was consistent with the input value.
\begin{table}
\caption{\label{tab:table2} Summary of systematic uncertainties in the measurement of $c\tau$ for $\Lambda_b$  and $B^0$ and their ratio. The
total uncertainties are determined by combining individual uncertainties in quadrature.}
\begin{ruledtabular}
\begin{tabular}{lccc}
Source & $\Lambda_{b}$ ($\mu$m) & $B^0$ ($\mu$m)  & Ratio \\
\hline
Alignment             & 5.4  & 5.4  &  0.002 \\
Distribution models   & 6.6  & 2.8  &  0.020 \\
Long-lived components & 6.0  & 13.6  &  0.022\\
Contamination         & 7.2  & $-$  &  0.016 \\
\hline
{Total}               & 12.7 & 14.9  &  0.034 \\
\end{tabular}
\end{ruledtabular}
\end{table}

The results of our measurement of the $\Lambda_b$ and $B^0$ lifetimes are summarized as:
\begin{eqnarray}
c\tau(\Lambda_b) = 365.1^{+39.1}_{-34.7} \mbox{ (stat)} \pm 12.7 \mbox{ (syst) } \mu\mbox{m,} \\
\nonumber c\tau(B^0) = 450.0^{+23.5}_{-22.1} \mbox{ (stat)} \pm 14.9 \mbox{ (syst) } \mu\mbox{m,}
\end{eqnarray}
\noindent from which we have:
\begin{eqnarray}
\tau(\Lambda_b) = 1.218^{+0.130}_{-0.115} \mbox{ (stat)} \pm 0.042 \mbox{ (syst) ps,}\\
\nonumber \tau(B^0) = 1.501^{+0.078}_{-0.074} \mbox{ (stat)} \pm 0.050 \mbox{ (syst) ps}.
\end{eqnarray}
\noindent These can be combined to determine the ratio of lifetimes:
\begin{equation}
\frac{\tau(\Lambda_{b})}{\tau(B^0)} = 0.811^{+0.096}_{-0.087} \mbox{ (stat)}\pm 0.034 \mbox{ (syst)},
\end{equation}
where we determine the systematic uncertainty on the ratio by calculating the ratio for each systematic source and
quoting the deviation in the ratio as the systematic uncertainty due
to that source. We combine all systematics in quadrature as shown in
Table~\ref{tab:table2}. The main contribution to the systematic
uncertainty of the lifetime ratio is due to the long-lived component
of the $B^0$ sample. This is expected since the $B^0$ is more likely
than the $\Lambda_{b}$ to be contaminated by mis-reconstructed $B$
mesons due to its lower mass. The ratio of lifetimes, using the
world average $B^0$ lifetime $\tau(B^0) = 1.527 \pm 0.008 $
ps~\cite{HFAG2006}, is
\begin{equation}
\frac{\tau(\Lambda_{b})}{\tau(B^0)} = 0.797^{+0.089}_{-0.080}.
\end{equation}

In conclusion, we have measured the $\Lambda_b$ lifetime in the fully reconstructed exclusive decay channel $J/\psi \Lambda$. The measurement is
consistent with the world average~\cite{HFAG2006}, and the ratio of $\Lambda_b$ to $B^0$ lifetimes is consistent with the most recent
theoretical predictions~\cite{Tarantino}.

%
We thank the staffs at Fermilab and collaborating institutions, 
and acknowledge support from the 
DOE and NSF (USA);
CEA and CNRS/IN2P3 (France);
FASI, Rosatom and RFBR (Russia);
CAPES, CNPq, FAPERJ, FAPESP and FUNDUNESP (Brazil);
DAE and DST (India);
Colciencias (Colombia);
CONACyT (Mexico);
KRF and KOSEF (Korea);
CONICET and UBACyT (Argentina);
FOM (The Netherlands);
Science and Technology Facilities Council (United Kingdom);
MSMT and GACR (Czech Republic);
CRC Program, CFI, NSERC and WestGrid Project (Canada);
BMBF and DFG (Germany);
SFI (Ireland);
The Swedish Research Council (Sweden);
CAS and CNSF (China);
Alexander von Humboldt Foundation;
and the Marie Curie Program.
%


\begin{thebibliography}{99}
\bibitem{etadef} $\eta=-\ln[\tan(\theta/2)]$, where $\theta$ is the polar angle with respect to the beamline.
\end{thebibliography}
\end{document}